\documentclass[a4paper]{jpconf}
\usepackage{graphicx}
\usepackage[utf-8]{inputenc}
\usepackage[square,sort&compress]{natbib}

\usepackage{aas_macros}

\begin{document}
\title{Star formation in the central 0.5~pc of the Milky Way}

\author{Thibaut Paumard}

\address{%
Laboratoire d'études spatiales et d'instrumentation en astrophysique (bât. 6)\\
Observatoire de Paris / CNRS\\
5 place Jules Janssen\\
F-92195 MEUDON CEDEX\\
FRANCE}

\ead{thibaut.paumard@obspm.fr}

\begin{abstract}
  The supermassive black hole candidate at the Galactic Center is
  surrounded by a parsec-scale star cluster, which contains a number
  of early type stars. The presence of such stars has been called a
  ``paradox of youth'' as star formation in the immediate vicinity of a
  supermassive black hole seemed difficult, as well as the transport
  of stars from far out in a massive-star lifetime. I will recall 30
  years of technological developments which led to the current
  understanding of the nuclear cluster stellar population.  The number
  of early type stars known at present is sufficient to access the 3D
  structure of this population and its dynamics, which in turn allows
  discriminating between the various possible origins proposed along
  the years.
\end{abstract}

\section{Introduction}
The Galactic Center (GC) is a unique laboratory for studying galactic
nuclei.  Given its proximity ($\simeq8$~kpc), processes in the GC can
be investigated at resolutions and detail that are not accessible in
any other galactic nucleus. The GC has many features that are thought
to occur in other nuclei (for reviews, see \citep{GenzelTownes1987,
  MorrisSerabyn1996, Mezgeretal1996, Alexander2005}).  It contains the
densest star cluster in the Milky Way intermixed with a bright
H~\textsc{ii} region (Sgr~A West or the Minispiral) and hot gas
radiating at X-rays. These central components are surrounded by a
$\simeq1.5$~pc ring/torus of dense molecular gas (the circum-nuclear
disk, CND). At the very center lies a very compact radio source,
Sgr~A*. The short orbital period of stars (in particular the B star
S2) in the central arcsecond around Sgr~A* show that the radio source
is a $3$--$4\times10^6\;\mathrm M_\odot$ black hole (BH) beyond any
reasonable doubt \cite{Schoedeletal2002,Ghezetal2003}.  The larger GC
region contains three remarkably rich clusters of young, high mass
stars: the Quintuplet, the Arches, as well as the parsec-scale cluster
around Sgr~A* itself.

I will first review the historical road leading to the discovery of
the nuclear cluster, which is paved with technological advances in the
fields of infrared high-resolution imaging and spectroscopy. Having
exposed the facts that are currently known about the nuclear cluster,
I will present the pending questions and the debate currently taking
place on the formation scenario for this cluster.

\section{History of Galactic Center Infrared Observations}
\subsection{The discovery of the massive star population}
In the early '70s, infrared instruments where limited to single-pixel
detectors. Two maps of the region, at 2.2 and 10~$\mu$m, were
published in 1975 \cite{BecklinNeugebauer1975}. Those maps were
obtained by scanning the region with an InSb detector and a $2.5"$
circular aperture, which represented a two-fold improvement over
previous work (e.g. \cite{RiekeLow1973}). These maps showed that the
infrared emission in the central parsec was concentrated in a few
compact sources. Those sources where numbered and referred to as,
e.g., ``source 7''. This denomination later evolved into ``infrared
source 7'', or ``IRS~7'' for short. These names are still widely used
in the literature. These sources can be found on the SIMBAD database
by using the less ambiguous acronym ``GCIRS~$n$'', for ``Galactic
Center Infrared Source''.

The discovery of the point-like radio source at the GC (later called
Sgr~A*) had just been reported the year before
\cite{BalickBrown1974}. Quite naturally, the positional agreement
between this source and one of the infrared compact source (IRS~16)
from Ref.~\cite{BecklinNeugebauer1975} was mentioned there, and the
authors suggested that they may be coincident with the position of
highest stellar density in the Galaxy. At the same time, it was
becoming apparent that the central parsec contained a
$\sim5\times10^6\;\mathrm M_\odot$ massive object. In 1977,
Ref.~\cite{Oort1977} discussed how natural it is to suspect that the
compact radio source ``might contain the engine providing the [...]
mass required''. Detection of a broad He~\textsc i emission line at
the location of IRS~16 was reported in 1982 \cite{Halletal1982}. The
authors tentatively interpreted their observation as the signature of
orbital motion of gas close to the putative black hole.

However, when the first two-dimensional infrared cameras became
available (e.g. \cite{ForrestPipher1986}), providing arcsecond
resolution, IRS~16 was resolved into several point sources, none of
which was coincident with the radio source. Additional spectroscopic
studies (e.g. \cite{AHH90}) have demonstrated that these sources were
normal stars: Wolf--Rayet stars (or close to that stage) to be
precise, fairly rare in the Galaxy, but not unique and more common in
the Magellanic Clouds. Wolf--Rayet stars are evolved, massive,
short-lived stars. The presence of many of them in the central parsec
(about 30 are currently known \cite{Paumardetal2006}) has been challenging to
understand: Ref.~\cite{Morris1993} demonstrates that the standard
scenario for star formation was severely inhibited in this
environment, mainly due to the very strong tidal forces from the black
hole.

\subsection{Characterizing the massive star population}
The first diffraction-limited images of the GC where obtained on the
ESO NTT (New Technology Telescope), a 3.5m-aperture facility located
at La Silla (Chile), using a speckle imaging technique. A few infrared
stars were detected in the central arcsecond around Sgr~A* in 1993
\cite{Eckartetal1995}, but no actual infrared counterpart to the radio
source was detected on these data. The same authors were rapidly able
to report the first proper motions of GC stars
\cite{EckartGenzel1996}, and of these faint stars in the central
arcsecond \cite{EckartGenzel1997}. Approximately at the same time, the
Keck telescope delivered its first images. Also using a speckle
imaging technique, it permitted detection of accelerations of the same
stars, providing an independent estimate of the enclosed mass
\cite{Ghezetal2000}.

In 2002, the adaptive-optics system NAOS coupled with the camera
CONICA (forming together NACO) was used on ESO VLT on the GC for the
first time, few months after the first light of NACO. Together with
the 8 years of NTT data obtained previously, these observations
provided the authors of Ref.~\cite{Schoedeletal2002} with a long
enough time baseline to constrain the orbit of S2 (a.k.a. S0-2, SIMBAD
identifier: [EG97] S2), one of these stars. Its orbital period,
$\simeq15.2$~yr, is short enough that, as of writing, it has already
been observed for more than one complete orbit.

In parallel to these dynamical studies, spectroscopic studies have
been conducted to unveil the nature of all these stars (both within
the central arcsecond and at larger radii). Spectro-imaging techniques
have been used to disentangle the spectra of the many stars in this
extremely dense environment (e.g. \cite{AHH90, Krabbeetal1991,
  Paumardetal2001}). Just like imaging studies, they have benefited
from improving spatial resolution, reaching the diffraction limit of
10m-class telescopes with SINFONI on ESO VLT and, more recently,
OSIRIS on Keck. The first stars detected in the early infrared studies
-- in particular the IRS~16 members -- were very bright blue
supergiants, akin to luminous blue variables (LBVs), presumably in the
transitional Ofpe/WN9 stage leading to the Wolf--Rayet state. Then,
even more evolved stars (WN and WC stars) have been found, filling the
upper part of the initial mass function (IMF). They are much less
luminous than the IRS~16 stars, because they have lost most of their
mass already, but exhibit very bright and broad metal emission
lines. In 2003, absorption lines were detected in S2 and the star was
classified B0V \cite{Ghezetal2003, Eisenhaueretal2003}. Finally, using
SINFONI, Ref.~\cite{Paumardetal2006} reported detection of many (about
20) main sequence OB stars, starting to fill the low mass end of the
IMF. The conclusions of this paper, in part disputed (notably by Lu et
al., see these proceedings), are representative of the state of the
art.

\section{Current knowledge}

Put together, all these studies allow a consistent picture to be
depicted. The parsec-scale nuclear cluster consists mainly of old
stars. This population is dynamically relaxed. The radiation in the
central parsec is however dominated by a younger population, which
formed during one (or perhaps two) short event(s) or weak starburst(s)
$6\pm2$~Myr ago.

\subsection{Stellar disks at the parsec scale}
These young stars orbit the black-hole in a coherent fashion. It was
already noted in 1996 from the radial velocities alone of $\simeq20$
emission line stars \cite{Genzeletal1996}. $\simeq 40$ stars are found
to lie on a fairly thin disk, rotating clockwise in projection on the
sky, called the clockwise system (CWS) \cite{Genzeletal2000,
  Genzeletal2003, Paumardetal2006, LevinBeloborodov2003,
  Tanneretal2006}. \cite{Genzeletal2003} and \cite{Paumardetal2006}
argue for the existence of a second disk (the counter-clockwise
system, CCWS) containing $\simeq15$ stars, using two different
methods. However, Lu et al. (these proceedings) have developed an
independent method for finding disks in the same data set as
\cite{Paumardetal2006} and testing their significance. They claim that
the CCWS found by their predecessors is not statistically
significant. Hendrik Bartko et al. (these proceedings) reacted to this
criticism, and took two actions to check their previous result: (1)
they also developed a more robust statistical approach to determine
the significance of their findings, and (2) they improved the quality
of their data set, by re-observing stars for which the radial
velocities were the least well constrained. They claim to confirm the
existence of the CCWS. Some more time is needed to allow the two
groups to converge to a common interpretation on this crucial
point. Although the existence of the CCWS is still controversial, I
give its proposed properties below.

While most stars on the CWS have fairly low eccentricities, the CCWS
seems to harbor more eccentric stars ($e\simeq0.8$). As the
measurement uncertainties decrease, it is also becoming apparent that
about half of the early type stars in the GC do not belong to either
disk. The two systems are fairly inclined with respect to the Galactic
plane and to one-another. They are both $6\pm 2$~Myr old, their
initial mass function is top heavy and their total mass is around
$10^4\;M_\odot$. The disk(s) do not extend down to Sgr~A*: they have
an inner radius of about 1 arcsecond, inside which the S star cluster
resides.

\subsection{A relaxed population in the Central arcsecond}
In contrast to the disk population, the S stars seem relaxed. They
have random eccentricities and orbital planes. There is no clear
indication so far of whether these stars were formed at the same time
as the disk stars, since none of the most evolved, WR stars belong to
the S cluster. The S stars could therefore be somewhat older than the
stellar disk population, but not much.

\subsection{IRS~13E: a cluster in the cluster}
Finally, there are strong claims about a very compact star cluster,
about $3"$ from Sgr~A* \cite{Maillardetal2004}. Although it has been
disputed \cite{Schoedeletal2005}, a conservative statistical analysis
has demonstrated that IRS~13E is a physical group of stars, containing
at least a dozen of stars, possibly many more
\cite{Paumardetal2006}. It is made of early-type stars of various
initial masses, including Wolf-Rayet stars. If the CCWS exists,
IRS~13E presumably belongs to it. It is bright at every wavelength
from the submillimetric domain to X-rays. It is a serious candidate
for harboring an intermediate mass black hole of
$10^{3-4}\;M_\odot$. However, the X-ray emission originates in the
colliding winds from the hot stars in the cluster rather than from an
accretion flow onto this putative black-hole. A conservative lower
limit on the cluster mass from its velocity dispersion is
$\simeq10^3$. However, the tidal forces from Sgr~A* impose another
lower limit to bound the cluster. Depending on its position on the
line-of-sight, it may be as high as $10^4\;M_\odot$. While it is
reasonable to assume the IRS~13E indeed contains $10^3\;M_\odot$ in
stars, a more massive cluster would require a compact, dark component.

\subsection{Ongoing star formation in the central parsec?}
So far, very little observations support the idea that there could be
current star formation in the central parsec. Just north of IRS~13E,
there is a group of very red point sources (IRS~13N), which have been
tentatively interpreted as young stellar objects \cite{Eckartetal2004,
  Muzicetal2008}, although with caution. Further investigations
are required to decide on this matter.

\section{Star formation scenarios}

\subsection{Adiabatic Cloud Collapse}
It has been demonstrated that the standard scenario for star
formation, adiabatic cloud collapse, is heavily suppressed in the
central parsec (see Ref.~\cite{Morris1993} for a thorough
discussion). Tidal forces are such that a cloud cannot be
gravitationally bound unless its density is in excess of
$10^7$~cm$^{-3}[1.6$~pc$/R_\mathrm{gc}]^{1.8}$ where $R_\mathrm{gc}$
is the distance to the Galactic nucleus \cite{GuestenDownes1980}. Such
densities are very hard to achieve, and require an external trigger,
for instance cloud collisions, strong winds, or supernova
shocks. Ref.~\cite{Morris1993} argues that the magnetic field in the
GC clouds has milligauss strength. In the central parsec this is
supported by polarization observations \cite{Aitkenetal1986,
  Aitkenetal1991}. Therefore, magnetic pressure is probably sufficient
to support these clouds.

\subsection{In-falling cluster scenario}
If producing stars seems very difficult in the central parsec, it
makes sense to try and externalize the star forming region. This
alternative in-falling star cluster scenario speculates that the GC
early-type stars were formed as a massive star cluster reasonably far
from Sgr~A*, were tidal forces are not so much an issue, and then
migrated towards their present location, loosing angular momentum
through dynamical friction. Several authors investigated this
possibility \cite{Gerhard2001, McMillanPortegiesZwart2003,
  PortegiesZwartetal2003, KimMorris2003, Kimetal2004,
  GurkanRasio2005}. Their collective conclusion is that this scenario
can work, but requires a very high initial mass ($>10^5\;M_\odot$) and
a very dense core ($>10^7\;M_\odot$pc$^{-3}$). When the core density
reaches this level, star--star collisions become likely and can lead
to runaway formation of an intermediate mass black-hole, which helps
the in-fall of the cluster towards the central parsec. During the
in-fall, the cluster undergoes mass-segregation and looses many
lower-mass stars. These elements provide a few predictions which can
be checked against the observations:
\begin{itemize}
\item residual core: since the cluster core ends up very tight, it may
  be able to survive for a long time in the central parsec. IRS~13E
  provides a natural candidate for having originated as the cluster
  core;
\item total mass: the observed cluster contains only
  $\simeq10^4\;M_\odot$, an order of magnitude below the predictions;
\item strong mass segregation and low mass stars lost at large radii:
  there is currently no indication of mass segregation in the
  early-type nuclear cluster. However, the statistics for OB stars at
  large ($>0.5$~pc) are still low, and no lower-mass star
  (e.g. spectral type later than A) has been detected so far.
\end{itemize}
Overall, this scenario works qualitatively rather reasonably but so
far fails to reproduce the facts quantitatively.

\subsection{Self-gravitating disk scenario}
While outsourcing star formation did not work quite as well as hoped,
other authors worked at improving the \textit{in situ} scenario. It
has been demonstrated earlier that an initially spherical cloud in the
central parsec will not be able to contract adiabatically under its
own gravity. Instead, it will orbit Sgr~A* and be sheared by the tidal
forces. After the material on the shortest orbits completes one orbit,
it will collide with the slowest parts of the cloud, forming a
dispersion ring \cite{Sanders1998}. This shock in itself has been
proposed as being able to compress the material enough to form stars,
but it seems difficult to form $10^4\;M_\odot$ of stars with this
localized trigger. Instead, the ring will evolve into a disk by
getting more circular, flatter, geometrically thinner, and
denser. More clouds may then come into the central region and merge
with this disk, which can thus build-up a considerable mass. A disk
configuration provides confinement where shearing cannot act against
density anymore. Circularization means that neighboring particles
acquire similar velocities. When the disk reaches a critical density,
the disk can become self-gravitating. At that point, numerical
simulations show that star formation proceeds very quickly and
efficiently. Most of the gas can actually participate in star
formation, which leaves very little material for accretion onto the
black hole: this actually is a problem for other galactic nuclei where
starbursts cohabit with a luminous black-hole. This is not an issue in
the GC though, since Sgr~A* is known to be very
underluminous. \cite{LevinBeloborodov2003, MilosavljevicLoeb2004}

\section{Open question: Forming the S stars}
There is one issue that none of the star formation scenarios proposed
above addresses satisfactorily: the presence of the S stars in the
innermost arcsecond. For them, the ``paradox of youth'' is even more
pronounced. Forming stars so close to the black hole is deemed nearly
impossible. Two alternative approaches have been studied to explain
the observations: rejuvenation of old stars and capture of eccentric
stars.

\subsection{Rejuvenation of young stars}
The basic idea here is that the blue, hot stars in the central
arcsecond are not what they seem. They really are old stars that have
had plenty of time to sink into the depth of the potential well. They
look young because something happened to them which makes their
photosphere take the size and temperature of a massive
star's:
\begin{itemize}
\item envelope stripping: the S stars would be red giants whose envelope
  has been ripped off during a close encounter with Sgr~A*, only their
  hot core remaining \cite{Alexander1999, Drayetal2006};
\item tidal heating: Ref.~\cite{AlexanderMorris2003} proposes that
  there exist squeezars, stars on very eccentric orbits around Sgr~A*
  which are heated by tidal interactions with the massive black hole;
\item mergers: the S-stars could be the product of several
  constructive collisions of low-mass stars, building up mass to that
  of a B star \cite{Genzeletal2003}.
\end{itemize}

All these scenarios seem to fail at the quantitative level. In
particular, since they have been proposed, precise spectroscopic
information has been obtained for the S stars. It is fairly easy to
produce the right color for the stars since they are in the
Rayleigh--Jeans regime at K-band. On the contrary, it seems quite
unnatural that such exotic scenarios should produce so normal-looking
stars.

\subsection{Capture of stars on eccentric orbits}
The other idea is that the S-stars really are normal B stars. They
have been formed further out, and moved into the central
arcsecond. The scenarios proposed above to bring the entire central
parsec early-type star population are hardly able produce the S-stars,
although Ref.~\cite{HansenMilosavljevic2003} claim that it is possible
using an in-falling star cluster with intermediate mass black hole,
and resorting to moderate envelope stripping to explain the lack of
Wolf-Rayet stars in the innermost region. An additional process helps
in bringing some B stars from the central few parsecs to the central
arcsecond. Overall, this class of scenarios uses three ingredients:
\begin{itemize}
\item a reservoir of B stars at reasonable distance from Sgr~A*;
\item a process to put them on very eccentric orbits, so that they
  pass close to Sgr~A*, crossing the central arcsecond;
\item a process to trap the very eccentric, B stars in the central
  arcsecond.
\end{itemize}

In Ref.~ \cite{GouldQuillen2003}, the B star is hypothesized to
initially be the lower mass component of a binary system. The trapping
process is exchange capture: when the binary comes close to Sgr~A* on
an eccentric orbit, the higher mass component ($\simeq60\;M_\odot$) is
ejected while the lower mass star remains on a tightly bound orbit
around the supermassive black hole. However, this process is very
inefficient: perhaps only $\simeq0.5\%$ of massive stars passing
within $\simeq130$~AU will inject lighter companions into S-star-like
orbits. Although the paper concentrates mostly on the trapping
process, it does provide a proposal of a reservoir: the binaries would
come from in-falling star clusters as described in the preceding
section. However, the authors have to speculate that it is reasonable
for such clusters to form on eccentric orbits.

In Ref.~\cite{AlexanderLivio2004}, the trapping process is
reversed. Here, the basic idea is that Sgr~A* is surrounded by a
cluster of stellar mass black holes. Each time a star of similar mass
passes through this cluster, there is a chance that the stellar mass
black hole is ejected and replaced by the star. They model the
reservoir as ongoing star formation at the parsec scale, which in
itself does not match the observations beyond the fact that there has
been indeed star formation recently. They also have to remains
somewhat speculative about how to put many B stars on very eccentric
orbits within their lifetime.

\section{Conclusion}
Over thirty years of technological developments in infrared astronomy
have allowed to resolve the nuclear star cluster at the GC to an
unequaled level. It is composed of a spheroidal component of late-type
stars and a organized population of early-type stars. Of order
$10^4\;M_\odot$ of stars have been formed $6\pm2$~Myr ago. They are
currently concentrated in one or perhaps two disks spanning the
central parsec. There is a tight cluster of early type stars, $3.5"$
from Sgr~A* in projection. There is no evidence of ongoing star
formation in the region. In addition, the central arcsecond contains
several B stars on random orbits.

Although it has been difficult to understand how massive, short-lived
stars could ever be observed so close to a supermassive black hole,
there exists now several scenarios to explain the parsec-scale disk(s)
of stars in an essentially satisfactory manner. However, the existence
of the S star cluster in the central arcsecond is still to be
considered a ``paradox of youth''.

We can already foresee that several types of observations could help
in resolving the current issues. First of all, continuing observations
with the current techniques will allow detecting accelerations for
more and more stars at larger and larger radii from Sgr~A*. This will
bring new insight to the three-dimensional structure of the early-type
star population. It could confirm or rule out the second disk. It
could also make apparent older disks, starting to dissolve in the
relaxed, late-type star cluster.

Secondly, there are several ways in which the future generations of
instrumentation will improve the situation. The GRAVITY experiment
\cite{Eisenhaueretal2005}, approved by ESO as a second generation
instrument for the Very Large Telescope Interferometer, will be able
to put tighter constraints on the mass profile in the central
arcsecond \cite{Paumardetal2005}. This will be valuable inputs for the
exchange capture scenarios in the preceding section. A spectro-imager
with a large field-of-view on an extremely large telescope would allow
investigating the density profile of early-type stars at large radii
from Sgr~A*. This would allow testing the in-falling star cluster
scenario.

\bibliographystyle{iopart-num}
\bibliography{paumard}

\end{document}